\documentstyle[preprint,eqsecnum,prl,aps,epsf]{revtex}
\begin{document}
%%%%%%%%%%%%%%%% READ THIS %%%%%%%%%%%%%%%% READ THIS %%%%%%%%%%%%%%
%%%%%%%%%%%%%%%%%%%%%%%%%%%%%%%%%%%%%%%%%%%%%%%%%%%%%%%%%%%%%%%%%%%%%
% This paper has figures appended in  a second part as a uuencoded
% compressed tar file with instructions for unpacking.  They will be
%automatically
% included in the text if you have a functioning epsf.tex.
% If you don't have that macro package (available from hep-th), or don't
% have the figure files, COMMENT OUT THE FOLLOWING LINE:
\input epsf
% If you do not already have epsf.tex (it comes with the dvips driver),
% you can print out the postscript files separately.
% WARNING: there is more than one version of epsf.tex dated 18 Jul 1990
% some figures will produce errors unless you have the most recent
% version. note that the version of epsf.tex that comes with the
% NeXTstep 2.0/2.1 distribution is not an up-to-date version. you should
% get epsf.tex from hep-th if your dvips is up-to-date but your epsf.tex
% is not.
%
%%%%%%%%%%%%%%%%%%%%%%%%%%%%%%%%%%%%%%%%%%%%%%%%%%%%%%%%%%%%%%%%%%%%%%%%%%%

\draft

\title{An $SO(5)$ Symmetric Ladder}
\author{D. Scalapino$^{1}$, Shou-Cheng Zhang$^{2}$ and W. Hanke$^{3}$}
\address{
$^{1}$Department of Physics,
University of California,
Santa Barbara, CA 93106
}
\address{
$^{2}$Department of Physics,
Stanford University
Stanford, CA 94305
}
\address{$^{3}$Institut f\"{u}r Theoretische Physik, Am Hubland,
D-97074 W\"{u}rzburg, Federal Republic of Germany}
\date{\today}
\maketitle
\begin{abstract}
We construct an $SO(5)$ symmetric electron model on a two-chain ladder
with purely local interactions on a rung. The ground state phase diagram
of this model is determined in the strong-coupling limit. The
relationship between the spin-gap magnon mode of the spin-gap insulator
and the $\pi$ resonance mode of the $d$-wave pairing phase is discussed.
We also present the exact ground state for an $SO(5)$ superspin model.
\end{abstract}

\newpage

\section{Introduction}
\label{sec:I}

The cuperate two-leg ladder materials are characterized by strong
electronic correlation and a variety of competing ground states
\cite{DR96}.
Numerical and analytic calculations for Hubbard \cite{NSW96}
and $t-J$ models \cite{TTR96} have
shown that at half-filling, these two-leg ladder models exhibit a 
spin-gap insulating phase and that when the system is initially doped,
$d_{x^2-y^2}$-like pairing and CDW correlation can become dominant. At
higher doping the system is expected to behave as a one-dimensional
Luttinger liquid \cite{BF96}. Given the strongly interactive 
nature of these systems
and the delicate balance between their competing ground states, one
would like to have a more general framework for determining their
properties.  Recently it was suggested that one could capture some of
the basic low energy physics with models having an $SO(5)$ 
symmetry\cite{so5,demler}.  As
we will discuss, there is in fact a natural way to construct an $SO(5)$
symmetric model for a two-leg ladder which has only {\it local}
interactions on a rung.  In this work,
we will show how to construct such a
Hamiltonian and discuss its strong-coupling phase diagram and low lying
collective excitations. 

This work is partially motivated to use the ladder system as
a theoretical laboratory to check some ideas of the $SO(5)$
theory. Recently, two dimensional lattice electronic models with 
exact $SO(5)$ symmetry have been constructed by three groups
independently\cite{rabello,henley,burgess}. However, these models
all involve long range interactions.
The $SO(5)$ symmetric ladder models constructed in this work 
involve only local interactions on the rung and are therefore much
easier to visualize. There has been considerable progress in numerically
checking the approximate $SO(5)$ symmetry in the 2D $t-J$ and Hubbard 
model\cite{meixner,eder}. The locally $SO(5)$ symmetric ladder models
constructed in this work are simple to implement numerically. By 
systematically varying the parameter away from the $SO(5)$ symmetric
point, one can trace the evolution of the $SO(5)$ multiplet 
structure \cite{eder} and get a better sense of the nature
of the approximate $SO(5)$ symmetry. In this work, we find a continuous
quantum phase transition from the spin gap Mott insulator phase to the
$d$ wave superconducting phase, and we show that the spin gap magnon
mode of the Mott insulator evolves continuously into the $\pi$
resonance mode\cite{demler,meixner,eder,demler2} 
of the superconducting phase. These results also
shed some light on the nature of the $\pi$ resonance mode in the 
2D case. Perhaps one of the most important
questions in the $SO(5)$ theory concerns the origin of such a symmetry
in generic models. 
Recently, Shelton and Senechal \cite{shelton}
studied the problem of two coupled 1D Tomonaga-Luttinger chains and
concluded that  approximate $SO(5)$ symmetry 
can emerge in the low energy limit
of this model. Balents, Fisher and Lin\cite{lin}
have used a weak 
coupling RG method combined with abelian bosonization to show that 
a generic ladder model at half-filling flows to a manifold with 
$SO(5)$ symmetry 
(and in fact, to various phases characterized by higher 
symmetries). 
Arrigoni and one of us (WH) additionally included a next-nearest-neighbor
hopping $t'$, which explicitly breaks the symmetry between the 
bonding and antibonding bands and thus the $SO(5)$ symmetry even in
the non-interacting limit\cite{enrico}. In this case, the model still
flows to an $SO(5)$ symmetric (at least up to order $O(t'/t)^2$) 
effective action, provided the $SO(5)$ symmetry is redefined by
accounting for different single-particle renormalization factors
for the fermions on the two bands of the ladder system.
These results indicate that the $SO(5)$ symmetry can emerge 
as a result of RG flow, and it is therefore of interest to study the
low-energy physics of 
fixed point Hamiltonians which have exact $SO(5)$ symmetry.

This paper is structured as follows.
The formal construction of an $SO(5)$ symmetric two-leg ladder
Hamiltonian will be described in Section II. Following this, Section III
contains a discussion of the ground state phases in the strong coupling
limit. The collective modes are discussed in Section IV,
and Section V contains a summary of our results.  

Before going into the technical details of the subsequent sections,
we would like to conclude the introduction by explaining the basic
idea. On a given rung of the ladder,
there are two sites with $16$ electronic states, as depicted in
Fig.~1. These $16$ states can be classified into 4 different
groups, (the $E_0$, $E_1$, $E_2$ and $E_3$ groups in Fig.~1),
each transforming as irreducible multiplets under $SO(5)$ \cite{eder}.
If a local Hamiltonian on a given rung is $SO(5)$ symmetric,
states within a given multiplet must have the same energy. However,
simple visual inspection shows that the states in the $E_2$ and 
$E_3$ groups are already degenerate for generic interactions respecting
spin invariance, particle-hole symmetry with respect to half-filling
and symmetry under interchanging two sites. Therefore, only
one condition is required to make the spin triplet magnon state
and the two ``pair states" in the $E_1$ group degenerate.
This condition turns out to be $J=4(U+V)$, relating the 
onsite interaction $U$, a near-neighbor interaction
$V$, and an spin exchange interaction $J$ on a rung.

\section{Construction of $SO(5)$ Symmetric Ladder Models}
\label{sec:II}
Rabello et al.~\cite{rabello} showed that the crucial step in constructing 
general exact $SO(5)$ symmetric models is to identify a 4 component 
fermion operator
which transforms according to the fundamental spinor irreps of
$SO(5)$, or its equivalent $Sp(4)$. 
The important point here is that the geometry of the 2-leg
ladder makes it natural to group fermion operators on a rung into a
4-component spinor, which leads to a Hamiltonian with purely local
interactions.  In the following we will use $x,y,..$ to denote
the position of a rung on a ladder, and $c_\sigma(x)$ and 
$d_\sigma(x)$
to denote the spin up ($\sigma=1$) and spin down ($\sigma=-1$)
fermion destruction operators on the upper and lower chain, respectively. Our
$SO(5)$ spinor operator is defined by
\begin{eqnarray}
\Psi_\alpha(x) = \left(
\begin{array}{c}
c_\sigma(x)\\
d^\dagger_\sigma(x)
\end{array}
\right)
\label{even-spinor}
\end{eqnarray}
for the  even rungs with $(-1)^x=1$, and
\begin{eqnarray}
\Psi_\alpha(x) = \left(
\begin{array}{c}
d_\sigma(x)\\
c^\dagger_\sigma(x)
\end{array}
\right)
\label{odd-spinor}
\end{eqnarray}
for the odd rung with $(-1)^x=-1$.
These spinor operators satisfy the canonical anticommutation relations:
\begin{eqnarray}
\{\Psi^\dagger_\alpha(x),\Psi_\beta(y)\}=\delta(x-y)
\delta_{\alpha\beta}
\label{twothree}
\end{eqnarray}
and
\begin{eqnarray}
\{\Psi_\alpha(x),\Psi_\beta(y)\}=
\{\Psi^\dagger_\alpha(x),\Psi^\dagger_\beta(y)\}=0
\label{twofour}
\end{eqnarray}
Using these spinor operators and the Dirac $\Gamma$ matrices (for
details see appendix A), we can construct a {\bf 5} dimensional
$SO(5)$ superspin vector
\begin{eqnarray}
n_a(x)=\frac{1}{2}\Psi^\dagger_\alpha(x) \Gamma^a_{\alpha\beta}
\Psi_\beta(x)
\label{twofive}
\end{eqnarray}
a {\bf 10} dimensional $SO(5)$ symmetry generator
\begin{eqnarray}
L_{ab}(x)=-\frac{1}{2}\Psi^\dagger_\alpha(x) \Gamma^{ab}_{\alpha\beta}
\Psi_\beta
(x)
\label{twosix}
\end{eqnarray}
and a {\bf 1} dimensional $SO(5)$ scalar
\begin{eqnarray}
\rho(x)=\frac{1}{2}\Psi^\dagger_\alpha(x) \Psi_\alpha(x)
\label{twoseven}
\end{eqnarray}
on a given rung $x$.
The {\it local} commutation relation between these operators are
given by
\begin{eqnarray}
 \left[ L_{ab},  L_{cd} \right] =
 - i( \delta_{ac} L_{bd} + \delta_{bd} L_{ac}
   - \delta_{ad} L_{bc} -\delta_{cc} L_{ad})
\label{twoeight}
\end{eqnarray}
\begin{eqnarray}
 \left[ L_{ab},  n_c \right] =
  -i( \delta_{ac} n_b - \delta_{bc} n_a )
\label{twonine}
\end{eqnarray}
and
\begin{eqnarray}
 \left[ L_{ab},  \rho\right] = 0
\label{twoten}
\end{eqnarray}
The superspin vector $n_a$ is related to the AF and SC operators
by
\begin{eqnarray}
 n_1       &=& \frac{\left(\Delta^{\dagger} + \Delta\right)}{2}
= \frac{1}{2} (-i c^\dagger \sigma_y d^\dagger + h.c.)  \\
\label{twoeleven}
 n_{2,3,4} &=& N_{x,y,z} 
= \frac{1}{2} (c^\dagger \vec\sigma c - d^\dagger \vec\sigma d ) \\
\label{twotwelve}
 n_5       &=& \frac{\left(\Delta^{\dagger} - \Delta\right)}{2i}
=-\frac{1}{2} (c^\dagger \sigma_y d^\dagger + h.c.) 
\label{twothirteen}
\end{eqnarray}
where we supressed the spinor index on $c_\sigma$ and $d_\sigma$.
The symmetry generators $L_{ab}$ are expressed in terms of the rung
spin $S_\alpha=\frac{1}{2}
(c^\dagger \sigma_\alpha c + d^\dagger \sigma_\alpha d )$, 
charge $Q=\frac{1}{2}(c^\dagger c+d^\dagger d-2)$ 
and the $\pi_\alpha$ operators
\begin{equation}
\pi^\dagger_\alpha = -\frac{1}{2}c^\dagger \sigma_\alpha \sigma_y d^\dagger
\label{twofourteen}
\end{equation}
with
\begin{eqnarray}
L_{ab} & = & \left(
\begin{array}{ccccc}
0 & \ & \ & \ & \ \\
\pi^\dagger_x\!+\!\pi_x & 0 & \ &\  & \ \\
\pi^\dagger_y\!+\!\pi_y & -S_z & 0 & \ & \ \\
\pi^\dagger_z\!+\!\pi_z & S_y & -S_x & 0 & \ \\
Q & \frac{1}{i}(\pi^\dagger_x\!-\!\pi_x )  &
\frac{1}{i}(\pi^\dagger_y\!-\!\pi_y
 ) &
    \frac{1}{i}(\pi^\dagger_z\!-\!\pi_z ) & 0
\end{array}
\right)
\label{twofifteen}
\end{eqnarray}
The $SO(5)$ singlet
 operator 
$\rho=\frac{1}{2}(c^\dagger c-d^\dagger d+2)$ 
has the physical interpretation of
the charge-density-wave operator.

Having exhibited the local $SO(5)$ operator algebra, we are now in a
position to construct an $SO(5)$ symmetric model. Let us first consider
the problem of two sites on a given rung. As discussed in the
introduction, there are 16 states which  
can be classified under $SO(5)$ as 1) a spin singlet state on the
rung 
\begin{eqnarray}
|\Omega\rangle = \frac{\left(c^\dagger_{\uparrow}
d^\dagger_{\downarrow} - c^\dagger_{\downarrow}
d^\dagger_{\uparrow}\right)}{\sqrt{2}} |0\rangle
\label{omega}
\end{eqnarray}
which is also an $SO(5)$ singlet ($L_{ab}|\Omega\rangle=0$),
2) an $SO(5)$ vector
quintet $n_a|\Omega\rangle$ which contains a triplet of ``magnon states"
and a doublet consisting of a hole pair and its conjugate, 3) two $SO(5)$
spinor quartets $\Psi_\alpha|\Omega\rangle$ and $\Psi^\dagger_\alpha|\Omega\rangle$,
and finally 4) two additional $SO(5)$ singlets of the form
$\Psi_\alpha R_{\alpha\beta} \Psi_\beta |\Omega\rangle$ and
$\Psi^\dagger_\alpha R_{\alpha\beta} \Psi^\dagger_\beta |\Omega\rangle$.
The $R$ matrix is a {\it invariant tensor} of the $SO(5)$ algebra and
it is defined in the Appendix A.
These states are depicted in Fig.~1.

Let us first neglect the hopping within the rung, and consider the
general spin and charge interaction Hamiltonian for the two sites:
\begin{eqnarray}
H_{rung} & = & U(n_{a\uparrow}-\frac{1}{2})(n_{a\downarrow}-\frac{1}{2})
+(a\rightarrow b) \nonumber\\
  & + & V (n_a-1)(n_b-1) + J \vec S_a \vec S_b
\label{rung-H}
\end{eqnarray}
In order for such a Hamiltonian to be $SO(5)$ symmetric, we would
require
that the states within each of the four multiplets mentioned above
be degenerate. As noted earlier, degeneracy within the
$\Psi_\alpha|\Omega\rangle$, $\Psi^\dagger_\alpha|\Omega\rangle$,
multiplets
is automatically ensured by spin rotation invariance,
the particle-hole symmetry and the
symmetry under rung parity $(c_\sigma\rightarrow d_\sigma)$. 
The $\Psi_\alpha R_{\alpha\beta} \Psi_\beta |\Omega\rangle$ and
$\Psi^\dagger_\alpha R_{\alpha\beta} \Psi^\dagger_\beta |\Omega\rangle$
states are already $SO(5)$ singlets. This leaves
only the $n_a|\Omega\rangle$ quintet manifold. For the rung Hamiltonian,
Eq (\ref{rung-H}), the triplet ``magnon states'' are seen to have 
energy $J/4-U/2$,
while the doublet pair states have energy $U/2 + V$. Therefore,
 these states will be
degenerate if 
\begin{equation}
J = 4(U+V)
\label{twoeighteen}
\end{equation}
Under this condition, the Hamiltonian (\ref{rung-H}) can be cast
into the manifestly $SO(5)$ symmetric form:
\begin{eqnarray}
H_{rung} = \frac{J}{4} \sum_{a<b} L_{ab}^2 + (\frac{J}{8}+\frac{U}{2})
(\Psi^\dagger_\alpha\Psi_\alpha-2)^2
\label{rung-so5}
\end{eqnarray}
up to an additive constant $3J/4+U/2$. For this $SO(5)$ symmetric form
of the interaction, 
the above mentioned four multiplets have the
energies $E_0 = -\frac{7}{2} U-3V$, $E_1 = U/2 + V$,
$E_2 = 0$ and $E_3 =  U/2 - V$ respectively as indicated in Fig.~1.

In the manifestly $SO(5)$ symmetric Hamiltonian (\ref{rung-so5}),
the interactions are expressed in terms of tensor and scalar
interactions. The readers may wonder why the vector interactions
are missing. The reason is that there is a Fierz identity
discussed in Appendix A, which relates the three channels of
interactions, and only two of them are mutually independent.

Let us now consider the effect of the hopping within the rung.
It is easy to see that the hopping term can be expressed in the
manifestly $SO(5)$ symmetric form:
\begin{eqnarray}
H_{t_\bot} & = & -2t_\bot (c^\dagger_\sigma d_\sigma + h.c)
 \nonumber\\
 & = & t_\bot (\Psi_\alpha R^{\alpha\beta} \Psi_\beta + h.c.)
\label{rung-hop}
\end{eqnarray}
This hopping term can split the degeneracy within the $E_2$ and
$E_3$ manifold. The $E_2$ manifold splits into anti-bonding and
bonding states
$(\Psi_\alpha\pm (R\Psi^\dagger)_\alpha)|\Omega\rangle$,
with energies $E_2^a=2t_\bot$ and $E_2^b=-2t_\bot$ respectively.
$H_{t_\bot}$ can in principle cause mixing between the $E_0$ and
the two states in the $E_3$ manifolds. Since all three states are
$SO(5)$ singlets, their mixing does not violate $SO(5)$ symmetry.
Within degenerate state perturbation theory, the two $E_3$ states
also form anti-bonding and bonding combinations
$(\Psi_\alpha R_{\alpha\beta} \Psi_\beta \pm
\Psi^\dagger_\alpha R_{\alpha\beta} \Psi^\dagger_\beta)|\Omega\rangle$,
with energies $E_3^a= U/2 - V-t^2_\perp/|2U+V|$ and $E_3^b= U/2-V +
t^2_\perp/|2U+V|$ respectively.

So far the $SO(5)$ symmetry is only realized on the two sites of
a rung. The more non-trivial question is how the symmetry is realized
when the hopping $t_\parallel$ in the ladder direction is included.
Remarkably,
this hopping can also be expressed in manifestly $SO(5)$ invariant
form,
\begin{eqnarray}
H_{t_\parallel} & = & -2t_\parallel \sum_{<x,y>} (c^\dagger_\sigma(x)
c_\sigma(y
)
+ d^\dagger_\sigma(x) d_\sigma(y)  + h.c) \nonumber\\
 & = & 2t_\parallel \sum_{<x,y>} (\Psi_\alpha(x) R^{\alpha\beta}
\Psi_\beta(y) +
 h.c.)
\label{ladder-hop}
\end{eqnarray}
It is important to point out that the alternating definition
of the $SO(5)$ spinors on the even (\ref{even-spinor}) and 
odd (\ref{odd-spinor}) rungs makes it
possible to express $H_{t_\parallel}$ in {\it manifestly} $SO(5)$
invariant form. This alternating definition was suggested to us
by S. Rabello. Without such an alternating definition, the resulting
Hamiltonian is still $SO(5)$ symmetric, but the symmetry is not
manifest.
Our final $SO(5)$ symmetric ladder Hamiltonian $H$ is given
by
the sum of (\ref{rung-so5}), (\ref{rung-hop}) and (\ref{ladder-hop}).
For this Hamiltonian, all ten $SO(5)$ generators
\begin{eqnarray}
L_{ab} = \sum_x L_{ab} (x)
\label{twotwentytwo}
\end{eqnarray}
are exactly conserved,
\begin{eqnarray}
\left[H,L_{ab}\right]=0
\label{twotwentythree}
\end{eqnarray}
Notice that because of our alternating definitions (\ref{even-spinor})
and (\ref{odd-spinor}) of the fermion operators,
the $\pi_\alpha$ operators have momentum $\pi$
along the ladder, while the total charge and total spin operators are
uniform. In the presence of a chemical potential term, $H_\mu=-2\mu
L_{15}$,
the $\pi_\alpha$ operators are exact eigen-operators
\begin{eqnarray}
\left[H+H_\mu,\pi^\dagger_\alpha\right]=-2\mu\pi^\dagger_\alpha
\label{eigenoperator}
\end{eqnarray}
so that the total Casimir charge of the ladder $C=\sum_{x,a<b} L_{ab}^2$
is conserved.
\begin{eqnarray}
\left[H+H_\mu,C\right]=0
\label{twotwentyfive}
\end{eqnarray}
Therefore, all states of the {\it doped} ladder are still labeled by
their
$SO(5)$ quantum numbers.

The $SO(5)$ symmetric ladder model we presented so far has only
local interactions on the rungs. Obviously, one can generalize the
model by including interaction between different rungs, for example
one could write down $SO(5)$ invariant interactions having the form 
\begin{eqnarray}
& & \sum_{x,y} V_1(x-y) n_a(x) n_a(y) + \sum_{x,y} V_2(x-y)
L_{ab}(x) L_{ab}(y)  \nonumber\\
& + & \sum_{x,y} V_0(x-y) (\rho(x)-2) (\rho(y)-2)
\label{twotwentysix}
\end{eqnarray}
Here we shall restrict ourselves only to the analysis of models
with local rung interactions, and defer the general analysis to
future works. It is plausible that in the strong coupling limit,
the local rung interaction dominates the physics.

\section{Strong Coupling Phase Diagram}
\label{sec:III}

In this section we discuss the phase diagram of the $SO(5)$ ladder
Hamiltonian in the strong coupling limit.  Setting $J=4(U+V)$, the
energies of the different rung manifolds are listed in Fig.~1. One can
divide up the $U$-$V$ plane according to regions in which a given
manifold lies lowest in energy and  Fig.~2 shows such a plot. In the strong
coupling regime, one can study a given sector of the $U$-$V$ plane,
using the virtual hopping processes due to $H_{t_\perp}$ and $H_{t_\Vert}$ to
resolve the degeneracies and determine the dynamics of the low lying
excited states.  In the following we examine the three different regions
$E_0$, $E_3$, and $E_1$ shown in the $U$-$V$ phase diagram of Fig.~2. 

\subsection{The $E_0$ Spin-Gap $d$-Wave Phase}
\label{subsec:III.A}

In
the region $E_0$ bounded by $V=-2U$ for negative values of $U$ and
$V=-U$ for positive values of $U$, the singlet rung state
$|\Omega\rangle$ (see Fig.~3a) 
is lowest in energy and the system is expected to be
in a spin-gap insulating ground state. In the strong-coupling limit this
state is simply a product of rung singlets.  The spin-gap corresponds to
the energy to create a magnon triplet, as illustrated in Fig.~3b, in
which a rung singlet is replaced by a magnon triplet from the $E_1$
manifold.  This costs an energy 
\begin{equation}
\Delta_{sg}=\left(\frac{U}{2} + V\right) + \left(\frac{7U}{2} + 3V\right) 
= 4(U+V) = J
\label{threeone}
\end{equation}

If we were to add two holes to this phase, the lowest energy state
occurs when the two holes are placed on the same rung as illustrated in
Fig.~3c. In this case, the excitation energy is again $\Delta_{sg}=J$, 
as expected
for an $SO(5)$ symmetric system.  Alternatively, one could imagine
adding two holes by placing each one on a separate rung creating two
$E_2=0$ states. However, this would cost an energy
$2(\frac{7}{2}\, U+3V)$, because of the two singlet rung states that are
destroyed, and this is a larger cost in energy than placing the holes on
the same rung throughout the entire $E_0$ region of Fig.~2. Thus the
doped holes will form rung pairs and we expect that the doped system will
exhibit power law pairing and CDW correlations in the $E_0$ region. If
one defines bonding and antibonding rung orbitals
\begin{equation}
b_\sigma^\dagger(x) = \frac{c_\sigma^\dagger (x) + d_\sigma^\dagger(x)}{\sqrt{2}}
\quad a_\sigma^\dagger(x)=\frac{c^\dagger_\sigma (x) - d_\sigma^\dagger(x)}{\sqrt{2}}
\label{threetwo}
\end{equation}
then the singlet rung state has the form
\begin{equation}
\left(\frac{c^\dagger_\downarrow d^\dagger_\downarrow - c^\dagger_\uparrow
d_\uparrow}{\sqrt{2}}\right)|0\rangle = \left(\frac{b^\dagger_\uparrow
b^\dagger_\downarrow - a^\dagger_\uparrow
a_\downarrow}{\sqrt{2}}\right)|0\rangle
\label{threethree}
\end{equation}
Thus in the $E_0$-region, the hole pairs go into a ``$d$-wave'' like
state \cite{TTR96} 
in which the amplitudes of the singlet pair in the bonding and
antibonding orbitals have opposite signs.

Both the magnon and the hole pair can propagate coherently along the
ladder leading to an energy dispersion in $q_x$.  In strong coupling we
can calculate their dispersion relations to second order in $t_\Vert$ as
follows.  A magnon excitation on rung $x$ can
hop to rung $x+1$ by going through an $E_1$ intermediate state. If
$|\psi_x\rangle$ is a state with the magnon on site $x$, then the
second order virtual hopping process has a matrix element
\begin{equation}
\langle\psi_{x+1} | H_{t_\Vert} \frac{1}{E_0-H_{\rm rung}} 
H_{t_\Vert} |\psi_x\rangle = \frac{t^2_\Vert}{3U+2V}
\label{threefour}
\end{equation}
and the magnon dispersion is
\begin{equation}
2\frac{t_\Vert^2}{3U+2V} \cos q_x
\label{threefive}
\end{equation}
There is also a $t^2_\Vert$ shift in the zero point energy when a magnon
is created. Taking these virtual processes into account gives the
complete magnon dispersion to second order in $t^2_\Vert$:
\begin{equation}
\omega_q = J\left(1-\frac{2t^2_\Vert}{(3U+2V)(\frac{7V}{2}+3V)}\right) +
\frac{2t^2_\Vert}{3U+2V} \cos q_x
\label{threesix}
\end{equation}

It is straight forward to carry out a similar calculation for the hole
pair dispersion and one finds that only the sign of the $\cos q_x$ term
in Eq.~(\ref{threesix}) is changed.  Thus, as expected for an $SO(5)$
symmetric ladder, the magnon dispersion about $q_x=\pi$ is identical to
the hole pair dispersion about $q_x=0$.

If two hole pairs (or two magnons) are added, one finds that they
repel each other if they are located on neighboring rungs. This
repulsion simply reflects the reduction in the zero point fluctuation
energy from the background fermions when the two hole pairs are adjacent. 
This gives rise to an effective near neighbor repulsion
$V^*= 4t^2_\Vert/(3U+2V)$. Therefore, in the $E_0$ region of the $U-V$
phase diagram the doped system behaves as a dilute one-dimensional
boson gas with a repulsive near-neighbor interaction, and is expected
show quasi-long-ranged-order in the superconducting correlation function.
The transition from the spin gap Mott insulator to the superconductor
occurs when the chemical potential is given by 
\begin{equation}
2|\mu|=2\mu_c= \Delta_{sg}
\label{threeseven}
\end{equation}
and such a transition is expected to be second order.

\subsection{The $E_3$ Spin-gap $s$-wave and CDW Phase}
\label{subsec:IIIB}

In the $E_3$-region of the $U$-$V$ strong coupling phase diagram (Fig. 2)
the singlet states 
$\Psi^\dagger_\alpha R_{\alpha\beta} \Psi^\dagger_\beta |\Omega\rangle$
and 
$\Psi_\alpha R_{\alpha\beta} \Psi_\beta |\Omega\rangle$ 
are coupled by second order kinetic energy processes. 
In this regime, the problem can be mapped onto an effective Ising
model in a magnetic field. We identify 
$\Psi^\dagger_\alpha R_{\alpha\beta} \Psi^\dagger_\beta |\Omega\rangle$
with the Ising state $|\sigma^z(x) = 1\rangle$ and 
$\Psi_\alpha R_{\alpha\beta} \Psi_\beta |\Omega\rangle$ 
with the Ising state $|\sigma^z(x) = -1\rangle$. 
On a rung we have
\begin{equation}
\langle \sigma^z(x) = 1 |H_{t_\perp} 
\frac{1}{E_0-H_0} H_{t_\perp} | \sigma^z(x) = -1 \rangle
= \frac{t^2_\perp}{2U+V}
\label{threeeight}
\end{equation}
and between two near neighbor rungs
\begin{equation}
\langle \sigma^z(x+1) = 1 | H_{t_\Vert}
\frac{1}{E_0-H_0} H_{t_\Vert} | \sigma^z(x) = -1 \rangle
= \frac{t^2_\Vert}{U/2-V}
\label{threenine}
\end{equation}

In the $E_3$-region, the energy denominators in the expressions are
negative so that the $t^2_\perp$ term favors the formation of the $s$-wave
like rung singlet
\begin{equation}
\Psi^\dagger_\alpha R_{\alpha\beta} \Psi^\dagger_\beta |\Omega\rangle
+ 
\Psi_\alpha R_{\alpha\beta} \Psi_\beta |\Omega\rangle
= \frac{b^\dagger_\uparrow
b^\dagger_\downarrow + a^\dagger_\uparrow
a^\dagger_\downarrow}{\sqrt{2}} |0\rangle
\label{threeten}
\end{equation}
Here $b^\dagger_\sigma$ and $a^\dagger_\sigma$ are the bonding and antibonding
creation operators of Eq (\ref{threetwo}). On the other hand, the
$t^2_\Vert$ process favors a staggered charge density wave state. 
Combining equations (\ref{threeseven}) and (\ref{threeeight}) 
we can write an effective Ising-like
Hamiltonian for the $E_3$-region in the form
\begin{equation}
{\cal H}_3 = \sum_x \left(-h \sigma^x(x) +
K_3 \bigl(\sigma^z_{x+1}\sigma^z(x) -1\bigr)\right)
\label{threeeleven}
\end{equation}
with $h=t^2_\perp/|2U+V|$ and $K_3=2t^2_{\Vert}/|U/2-V|$. The ground state
of ${\cal H}_3$ is known to have an Ising-like phase transition for $h=K_3$.
For $h< K_3$, the half-filled $SO(5)$ ladder will be in a CDW phase
corresponding to one of the two degenerate states illustrated in 
Fig.~4a and Fig.~4b.
For $h>K_3$, the system will be disordered.  In this region, 
the half-filled $SO(5)$
ladder will be in a spin-gap insulating phase. For $t_{\Vert}=t_\perp$, the
$h=K_3$ dividing line corresponds to $V=-\frac{3}{4} U$. When holes are
doped into the disordered region, they will tend to go onto a rung
forming $s$-wave like, Eq (\ref{threeten}), pairing correlations.

\subsection{The $E_1$ $SO(5)$ Superspin Phase}
\label{subsec:III.C.}

Here the Hilbert space per rung is restricted to the
``superspin" quintet manifold $n_a(x)|\Omega\rangle$.
In this case, each rung is either occupied by a triplet
magnon or a doublet ``pair" state. 
The effective Hamiltonian in the quintet manifold is
easily determined using second order perturbation theory:
\begin{eqnarray}
{\cal H}_1=K_1 \sum_{<x,y>} L_{ab}(x) L_{ab}(y)
\label{threetwelve}
\end{eqnarray}
where $K_1= t^2_{\parallel}/|U/2+V|$. This model can be
viewed as the $SO(5)$ generalization of the spin one Heisenberg chain.
Therefore, we would expect to find many properties simply from this
analogy,
for example, a ground state with a finite excitation gap, and short
 ranged correlations, etc.
One useful model of the spin one Heisenberg chain is the AKLT\cite{aklt}
model for which an exact ground state is known. In this section we shall
construct
an $SO(5)$ generalization of the AKLT model and present its exact
ground state.

We begin by considering two neighboring rungs $x$ and $y$. The
wave function for the two superspins defined on the two rungs
can be decomposed as
\begin{eqnarray}
{\bf 5} \times {\bf 5} = {\bf 1} + {\bf 10} + {\bf 14}
\label{threethirteen}
\end{eqnarray}
{\it i.e.} the product wave function can transform like an
$SO(5)$ singlet, an $SO(5)$ antisymmetric tensor or an
$SO(5)$ symmetric traceless tensor. Therefore, we can defined
a complete set of bond projection operators $P_{\bf 1}(xy)$,
$P_{\bf 10}(xy)$ and $P_{\bf 14}(xy)$ onto these subspaces, satisfying:
\begin{eqnarray}
1 = P_{\bf 1}(xy) + P_{\bf 10}(xy) + P_{\bf 14}(xy)
\label{identity}
\end{eqnarray}
The $SO(5)$ generalization of the AKLT model is then given by 
\begin{eqnarray}
{\tilde{\cal H}_1} = 2K_1 \sum_{<x,y>} P_{\bf 14}(xy)
\label{so5aklt}
\end{eqnarray}
Let us first see how this Hamiltonian can be expressed in terms 
of the $SO(5)$ superspin exchange operators. We start by
defining
\begin{eqnarray}
{\cal L}_{ab} = L_{ab}(x) + L_{ab}(y)
\label{threesixteen}
\end{eqnarray}
where ${\cal L}_{ab}$ measures the total $SO(5)$ generator on the
bond $<xy>$. Squaring this equation and noticing that the Casimir
charge $C_{\bf 5}= \sum_{a<b} L_{ab}^2(x)$ for the {\bf 5} irreps
on a given rung is $C_{\bf 5}=4$, we obtain
\begin{eqnarray}
\sum_{a<b} L_{ab}(x) L_{ab}(y) = \frac{1}{2} \sum_{a<b}
{\cal L}_{ab}^2 -4
\label{threeseventeen}
\end{eqnarray}
The operator $\sum_{a<b} {\cal L}_{ab}^2$ is the total Casimir charge
for the bond $<xy>$, therefore, it can be expressed as
\begin{eqnarray}
\sum_{a<b} {\cal L}_{ab}^2 = C_{\bf 1} P_{\bf 1}(xy)
+ C_{\bf 10} P_{\bf 10}(xy) + C_{\bf 14} P_{\bf 14}(xy)
\label{threeeighteen}
\end{eqnarray}
where $C_{\bf 1}=0$, $C_{\bf 10}=6$ and $C_{\bf 14}=10$ are the
Casimir charge for the {\bf 1}, {\bf 10} and {\bf 14} irreps
respectively. Therefore, we obtain
\begin{eqnarray}
\sum_{a<b} L_{ab}(x) L_{ab}(y) = -4 P_{\bf 1}(xy)
- P_{\bf 10}(xy) + P_{\bf 14}(xy)
\label{exchange}
\end{eqnarray}
Squaring this equation again gives
\begin{eqnarray}
(\sum_{a<b} L_{ab}(x) L_{ab}(y))^2 = 16 P_{\bf 1}(xy)
+ P_{\bf 10}(xy) + P_{\bf 14}(xy)
\label{exchange2}
\end{eqnarray}
where we used the property of the projection operators
$P_i P_j = \delta_{ij} P_i$. Equations (\ref{identity}),
(\ref{exchange}) and (\ref{exchange2}) finally allows us to express
the $P_{\bf 14}(xy)$ operator as
\begin{eqnarray}
P_{\bf 14}(xy) = \frac{1}{10} (L_{ab}(x) L_{ab}(y))^2 +
\frac {1}{2} (L_{ab}(x) L_{ab}(y)) + \frac{2}{5}
\label{threetwentyone}
\end{eqnarray}
Inserting this equation into (\ref{so5aklt}) then gives the
desired expression for the $SO(5)$ generalization of the AKLT
Hamiltonian.

The Hamiltonian (\ref{so5aklt}) has an exact ground state. In general,
the ground state can be expressed as
\begin{eqnarray}
|\psi_0\rangle=\sum_{a_1,..,a_N} \psi_0(a_1,..,a_N)
n_{a_1}(x_1)..n_{a_N}(x_N)|\Omega\rangle
\label{threetwentytwo}
\end{eqnarray}
for a ladder with $N$ rungs. $\psi_0(a_1,..,a_N)$ is the corresponding
ground state wave function in the superspin vector basis. It is easy
to see that the exact ground state wave function for the $SO(5)$ AKLT
Hamiltonian (\ref{so5aklt}) for a {\it period} ladder is given by
\begin{eqnarray}
\psi_0(a_1,a_2,..,a_N)
= Tr(\Gamma^{a_1}\Gamma^{a_2}...\Gamma^{a_N})
\label{threetwentythree}
\end{eqnarray}
This follows from the following property of the Dirac $\Gamma$ matrices
(for more details, see appendix A):
\begin{eqnarray}
\Gamma^{a}\Gamma^{b}= 2 \delta^{ab} + 2 i \Gamma^{ab}
\label{threetwentyfour}
\end{eqnarray}
{From} this equation we see that the product of two $\Gamma$ matrices
involves no symmetric traceless components. Therefore, the wave
function $\psi_0(a_1,..,a_i,a_{i+1},..,a_N)$ viewed as a $5\times 5$
matrix in $a_i$ and $a_{i+1}$ with all other indices fixed has
no symmetric traceless components. This means that the {\bf 14}
irreps on bond $<x_ix_{i+1}>$ are absent, which implies that
$|\psi_0\rangle$ is annihilated by the projector Hamiltonian
(\ref{so5aklt}). Since the Hamiltonian (\ref{so5aklt}) is positive
definite, we can conclude that $|\psi_0\rangle$ is indeed the exact
ground state. 

For a ladder with {\it open} boundary conditions, the corresponding
ground state wave function is given by
\begin{eqnarray}
\psi_0(a_1,a_2,..,a_N)
= \Gamma^{a_1}\Gamma^{a_2}...\Gamma^{a_N}
\label{threetwentyfive}
\end{eqnarray}
This implies $4$ edge states at each end of the ladder, giving rise
to a $16$ fold ground state degeneracy.

To our knowledge, this is the first exact solution
to a problem of interacting magnon and Cooper pairs.
The ground state described by (\ref{threetwentytwo}) and
(\ref{threetwentythree}) is translationally invariant in the
ladder direction and is an $SO(5)$ singlet state. It has short
ranged antiferromagnetic and superconducting order along the
ladder. Because it is a resonating state, it is hard to 
draw a simple picture for this state. The spin part of this 
wave function can be basically visualized as  resonating 
between the states depicted in Fig.~5a and Fig.~5b.

\section{Collective Modes}
\label{sec:IV}
As discussed in Section IIIA in the $E_0$-phase the magnon dispersion
relation about $q_x=\pi$ is the same as the one hole pair dispersion
relation around $q_x=0$. Here we examine what happens when the system 
becomes superconducting upon doping with a finite 
concentration of hole pairs. 

The rung operator $L_{15}(x)$ is equal to
\begin{equation}
L_{15}(x) = {1\over 2}\left(N_e(x)-2\right)
\label{fourone}
\end{equation}
with
\begin{equation}
N_e(x) = \sum_s \left(c^\dagger_s(x)c_s(x) + d^\dagger_s(x)
d_s(x)\right)
\label{fourtwo}
\end{equation}
Thus $Q=\sum_xL_{15}(x)$ counts the number of pairs relative to
half-filling and in the presense of a chemical potential $\mu$, as
discussed in Section II, one adds the term
\begin{equation}
H_\mu = - 2\mu Q
\label{fourthree}
\end{equation}
to the ladder Hamiltonian $H$.  In the $E_0$-phase, as $\mu$ becomes 
increasingly negative, the
$E_1$ quintet splits with the $\Delta|\Omega\rangle$ mode linearly
decreasing its energy, the $\Delta^\dagger|\Omega \rangle$ mode linearly
increasing its energy and the $\vec N|\Omega\rangle$ magnon
modes remaining constant in energy, as illustrated in Fig.~6a. 
When $2|\mu|$ becomes greater than the
spin gap $\Delta_{sg}$, the ladder becomes doped with a finite density
of hole pairs.  As
discussed in Section III, these hole pairs behave in strong coupling
like a dilute hard core bose system with a near neighbor repulsion. 
In one dimension, hard
core bosons can be treated
as spinless fermions. Therefore, one can imagine that this band
is filled with spinless fermions up to some Fermi energy,
and $2|\mu|$ can be physically identified with this Fermi energy. 
The physical origin of the increase in the energy to add an extra hole pair
is due to the hard-core repulsion with the hole pairs in the
condensate.

On the other hand, the magnon band has a band minimum at momentum
$\pi$ and it is completely empty. A naive argument would suggest that
one could insert a magnon simply by putting it at the band minimum which
would
cost energy $\Delta_{sg}$. However, this argument neglects the
repulsive interaction between the magnon and the hole pairs in the
condensate. In an $SO(5)$ model, the interaction between the
magnon and the hole pairs are the same as the mutual interaction
between the hole pairs, therefore, the energy of the magnon is
the same as the energy of a hole pair, which is
$2|\mu|$. Thus we arrive at a simple physical interpretation
of the energy of the magnon in the
superconducting state: {\it the magnon
energy $2|\mu|$ in the superconducting
state is the sum of two contributions, the rest energy $\Delta_{sg}$
to create a magnon and the interaction energy between the magnon and
the hole pair in the condensate}.
Therefore, the energy of the spin triplet momentum 
$\pi$ excitation is $\Delta_{sg}$ for $|\mu|<\mu_c$ and $2|\mu|$ for
$|\mu|>\mu_c$, see (Fig.~6b).

This physical interpretation was based upon a strong-coupling
picture, but a more general result can be obtained as follows.
According to Eq.~(\ref{eigenoperator}) for an $SO(5)$ ladder the $\pi$
operators are exact eigen-operators
\begin{equation}
\left[H+H_\mu, \pi_\alpha\right] = 2\mu\pi_\alpha
\label{fourfour}
\end{equation}
which add a pair and generate an exact excited state with momentum $(\pi,
\pi)$ and $S=1$. Now, suppose we were to start with the ground state
of $H$ with charge $Q$, $|\psi_0(Q)\rangle$, and add a hole pair to
obtain the ground state of the electron system with charge $Q-1$,
$|\psi_0(Q-1)\rangle$. The energy cost to insert the hole pair is
given by the difference in the ground state energy of $H$ for $Q-1$ and
$Q$ electron pairs,
$2|\mu|=E_0(Q-1) - E_0(Q)$. On the other hand, we can act on
$|\psi_0(Q-1)\rangle$ with the $\pi^\dagger_\alpha$ operator and rotate the
added hole pair into a magnon of the $Q$ electron pair system.
For the $SO(5)$ ladder, this rotation costs no energy, so we see
that the energy for inserting a magnon into the $Q$ electron pair system is
$2|\mu|$. Furthermore, one sees that the $\pi$-mode is just the natural
continuation of the spin gap magnon mode \cite{Ref1}.  This corresponds to the
idea of the $\pi$-mode in the two dimensional $t-J$ model
originally proposed by Demler {\it et al}\cite{demler,demler2} and studied in numerical
calculation by Meixner et.~al.~\cite{meixner} and Eder et.~al.~\cite{eder}. 

The triangular relation between the $\pi$ resonance, the magnon, and the
Cooper pair can be illustrated by the $SO(5)$
representation theory discussed by Eder et.~al \cite{eder}.
The general
traceless symmetric tensor irreps of $SO(5)$ are characterized by three
intergers $(S_z, Q, \nu)$, where $\nu$ is related to the Casimir charge
by $L^2_{ab} = \nu(\nu+3)$. This class of eigenstates of any $SO(5)$
model can therefore be represented by an $SO(5)$ pyramid, labeled by the
Cartesian coordinates $(S_z, Q, \nu)$. If we take an $S_z=0$ slice of
the $SO(5)$ pyramid, and plot the resulting energy diagram, we obtain
Fig.~7.  Each box in Fig.~7 denotes the collection of all eigenstates
with the same $SO(5)$ quantum numbers.  The superconducting states lie
on the ridge of the pyramid. In the canonical ensemble, all states with
the same $\nu$ are strictly degenerate, but the lowest energy states with
$\Delta\nu=1$ are spaced by $2|\mu| = E_0 (Q-1)-E_0(Q)$. Starting from
the ground state $|\psi_0(Q)\rangle$, one can create a magnon using the
$N_\alpha$ operator, or by first adding a hole pair using the $\Delta$
operator, and then acting on the resulting $|\psi_0(Q-1)\rangle$ state
with the $\pi^\dagger_\alpha$ operator. (See Fig.~7). In $SO(5)$
symmetric models, this triangle closes exactly, and the magnon mode
energy is therefore predicted to be exactly $2|\mu|$.

\section{Conclusion}
\label{sec:V}

We have found that a two-leg ladder with a rung interaction
characterized by an onsite $U$ interaction, a rung near neighbor $V$
interaction, and a rung exchange interaction $J$ can have $SO(5)$
symmetry if $J=4(U+V$). Furthermore, in the $E_0$ regime, the
strong-coupling ground state is a spin-gap insulator.  In this
half-filled state the equal time rung magnetization and rung pair field
correlations are identical. In addition, and of particular importance,
the dispersion relation of a magnon rung excitation with $q_x$ measured
from $\pi$ is identical to the rung hole pair dispersion measured from
$q_x=0$. We have also seen that when the chemical potential is increased
such that $2|\mu|$ exceeds the spin gap $\Delta_{sg}$, 
$d$-wave-like hole pairs form a
dilute hard core bose gas with a near neighbor repulsion. The spin
gap magnon mode of the Mott insulator evolves continuously into the
$\pi$ resonance mode of the superconductor.

There are also other ground states in the $U-V$ phase diagram such as
the $E_3$ regime which can have a CDW state or a spin gap insulating
phase which when doped has $s$-wave hole pairs. In addition, the
$E_1$-phase at half-filling corresponds to an $SO(5)$-like Heisenberg
model with ground state gaps analogous to the $S=1$ Heisenberg model.

A key question remains regarding the relationship of this $SO(5)$ ladder
to the more standard Hubbard or $t-J$ ladders. Physically if we want $J$
and $U$ to be positive, this requires a negative rung interaction $V$.
Furthermore $|V|$ must nearly balance $U$ in order for the system to be
in the physical interesting regime in which $J/t<1$. Therefore,
the standard ladder models are not exactly $SO(5)$ symmetric
in the sense defined in this paper. However, 
at half-filling, it is likely that
standard ladder models flow towards a rung singlet ground state
in the strong coupling limit.
In this work we showed that such a state is not only a total
spin singlet, but also an $SO(5)$ singlet. Therefore, we would
expect the static correlation to be approximately $SO(5)$ symmetric.  
Recent results by Shelton and Senechal \cite{shelton}, 
Balents, Fisher, and Lin\cite{lin} and Arrigoni and Hanke\cite{enrico}
show that
the generic interaction parameters of the 
ladder model tend to flow towards the $SO(5)$ symmetric manifold
under RG. However, their results were obtained in the weak coupling
regime. Clearly, there remain various questions such as whether one
will have a sufficient renormalization flow to approach the $SO(5)$
regime when the physical $U$ is of order the bandwidth. 
The $SO(5)$ symmetric model studied in this work offer a reference
point around which departures from the $SO(5)$ symmetric point
can be studied systematically. It would be desirable to develop
a numerical RG analysis to study the flow around
the $SO(5)$ symmetric point in the strong coupling limit. 

\section*{Acknowledgments}
We would like to thank Dr. S. Rabello for a helpful suggestion in 
defining the alternating form of a 
$SO(5)$ spinor on the ladder and Dr. H. Kohno for
providing us with some useful identities in the Appendix.
We would like to acknowledge useful discussions with L.~Balents,
E. Demler, M.P.A.~Fisher, W. Kohn, H.-H.~Lin, A.~Millis,
and S.R.~White.
SCZ is supported in part by the NSF under grant numbers DMR94-00372
and DMR95-22915. DJS acknowledges support from NSF under grant number
DMR95-27304, and WH is supported by FORSUPRA II, BMBF (05 605 WWA 6), ERB
CHRXCT940438.

\appendix
\section{Dirac $\Gamma$ matrices and Fierz identity}
The general method introduced by Rabello et.~al to construct $SO(5)$
symmetric models uses the five Dirac $\Gamma$ matrices 
$\Gamma_a$ $(a=1,..,5)$ which satisfy the
Clifford algebra,
\begin{eqnarray}
\{\Gamma^a, \Gamma^b \} = 2 \delta^{ab}
\end{eqnarray}
Rabello et.~al introduced the following explicit representation which is
naturally adapted for discussing the unification of AF and dSC
order parameters,
\begin{eqnarray}
 \Gamma^1\!=\! \left( \begin{array}{cc}
               0          & -i\sigma_y  \\
               i\sigma_y  & 0            \end{array} \right)
 \Gamma^{(2,3,4)} \!=\! \left( \begin{array}{cc}
             \vec \sigma  & 0  \\
                  0       &  ^t\vec \sigma  \end{array} \right)
 \Gamma^5 \!=\! \left( \begin{array}{cc}
                0         & \sigma_y  \\
                \sigma_y  & 0            \end{array} \right)
\end{eqnarray}
Here $\vec \sigma=(\sigma_x,\sigma_y,\sigma_z)$ are the usual
Pauli matrices and $^t\vec \sigma$ denotes their transposition.
These five $\Gamma_a$ matrices form the $5$ dimensional vector
irreps of $SO(5)$. Their commutators
\begin{eqnarray}
\Gamma^{ab} = - \frac{i}{2} \left[ \Gamma^a, \Gamma^b \right]
 \end{eqnarray}
define the $10$ dimensional antisymmetric tensor irreps of
$SO(5)$. In the above representation, the $10$ $\Gamma^{ab}$'s are given
explicitly by
\begin{eqnarray}
 \Gamma^{15} &=&  \left( \begin{array}{cc}
                      -1  & 0  \\
                       0  & 1           \end{array} \right)
\nonumber \\
 \Gamma^{(i+1)(j+1)} &=&  \varepsilon_{ijk}\left( \begin{array}{cc}
                \sigma_k  & 0  \\
                     0    &  -{}^t\sigma_k  \end{array} \right)
 {} \hskip 7mm (i,j=1,2,3)
\nonumber \\
 \Gamma^{(2,3,4) 1} &=&  \left( \begin{array}{cc}
                0         & - \vec \sigma \sigma_y  \\
            - \sigma_y \vec \sigma & 0    \end{array} \right)
                    =  \sigma_y \left( \begin{array}{cc}
                0         & {}^t\vec \sigma  \\
            - \vec \sigma & 0    \end{array} \right)
\nonumber \\
 \Gamma^{(2,3,4) 5} &=&  \left( \begin{array}{cc}
                0         & -i \vec \sigma \sigma_y  \\
             i\sigma_y \vec \sigma & 0    \end{array} \right)
                    =  i\sigma_y \left( \begin{array}{cc}
                0         & {}^t \vec \sigma   \\
              \vec \sigma &  0    \end{array} \right)
\nonumber
\end{eqnarray}
These $\Gamma$ matrices satisfy the following commutation relations:
\begin{eqnarray}
 \left[ \Gamma^{ab}, \Gamma^c \right] &=& 2i(\delta_{ac}\Gamma^b -
                                  \delta_{bc}\Gamma^a)    \\
 \left[ \Gamma^{ab}, \Gamma^{cd} \right] &=&
          2i(\delta_{ac}\Gamma^{bd} + \delta_{bd}\Gamma^{ac}
           - \delta_{ad}\Gamma^{bc} - \delta_{bc}\Gamma^{ad})
\end{eqnarray}

A very important property of the $SO(5)$ Lie algebra is the
pseudo-reality of its spinor representation. This means that
there exists a matrix $R$ with the following properties:
\begin{eqnarray}
  R^2 = -1, \ \ \ R^{\dagger} =  R^{-1} = {}^tR = -R   \\
  R\,\Gamma^aR = -{}^t\Gamma^a, \ \ \ R\,\Gamma^{ab}R = {}^t\Gamma^{ab}
\end{eqnarray}
The relations
$ R\,\Gamma^{ab}R^{-1} = -(\Gamma^{ab})^*$ indicate that the spinor
representation is real, and the antisymmetric nature of the matrix $R$
indicates that it is pseudo-real. The $R$ matrix plays a role similar
to that of $\epsilon_{\alpha\beta}$ in $SO(3)$. In our representation,
the $R$ matrix takes the form
\begin{eqnarray}
R = \left( \begin{array}{cc}
                     0  &  1  \\
                    -1  &  0  \end{array} \right)
\end{eqnarray}

The sixteen $\Gamma^A=1, \Gamma^a, \Gamma^{ab}$ matrices form a
complete basis in the space of $4\times 4$ Hermitian matrices.
This basis is orthonormal by virtue of the trace operation:
\begin{eqnarray}
Tr(\Gamma^A \Gamma^B) = 4 \delta^{AB}
\end{eqnarray}
Therefore, any $4\times 4$ Hermitian matrix ${\bf M}_{\alpha\beta}$
can be expanded as
\begin{eqnarray}
{\bf M}_{\alpha\beta} = \sum_A \lambda_A \Gamma^A_{\alpha\beta}
\end{eqnarray}
with
\begin{eqnarray}
\lambda_A = \frac{1}{4} Tr ({\bf M} \Gamma^A)
\end{eqnarray}

This observation can be used to derive a series of Fierz identities,
relating interactions in the scalar, vector and tensor channels.
For example, the fermion bilinear
$\Psi^\dagger_\alpha(x) \Psi_\beta(y)$ can be expanded as
\begin{eqnarray}
\Psi^\dagger_\alpha(x) \Psi_\beta(x)
= \frac{1}{4} \sum_A (\Psi^\dagger(x) \Gamma^A \Psi(y))
\Gamma^A_{\beta\alpha}
\end{eqnarray}
Using this Fierz identity, one can show that the scalar, vector and the
tensor interactions {\it on the same rung} are not independent of each
other. They are instead related by the following equation
\begin{eqnarray}
(\Psi^\dagger \Psi -2)^2
= 4 - \frac{1}{5} (\Psi^\dagger \Gamma^a \Psi)^2 -
\frac{1}{5} (\Psi^\dagger \Gamma^{ab} \Psi)^2
\end{eqnarray}

\begin{figure*}[h]
\centerline{\epsfysize=8cm
\epsfbox{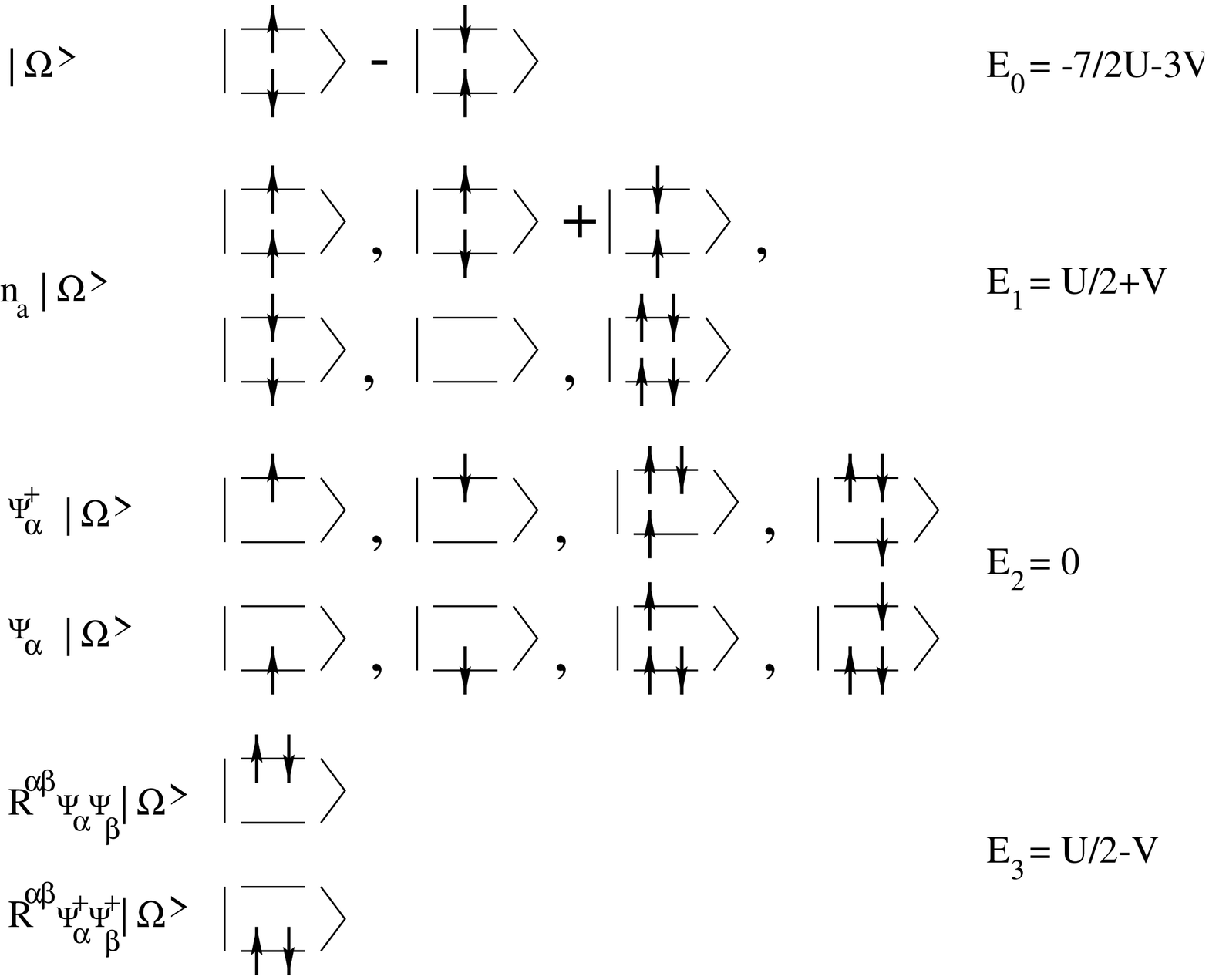}
}
\nonumber
\vskip .13 in
\caption{The 16 states of a rung are laid out in their $SO(5)$
multiplets. Here the two lines in the kets represent the two sites of a
rung which can be in four states (empty, one electron spin up, spin
down, or two electrons). The energies of the multiples for $J=4(U+V)$
are also listed.
}
\end{figure*}

\newpage

\begin{figure}[h]
\centerline{\epsfysize=8cm
\epsfbox{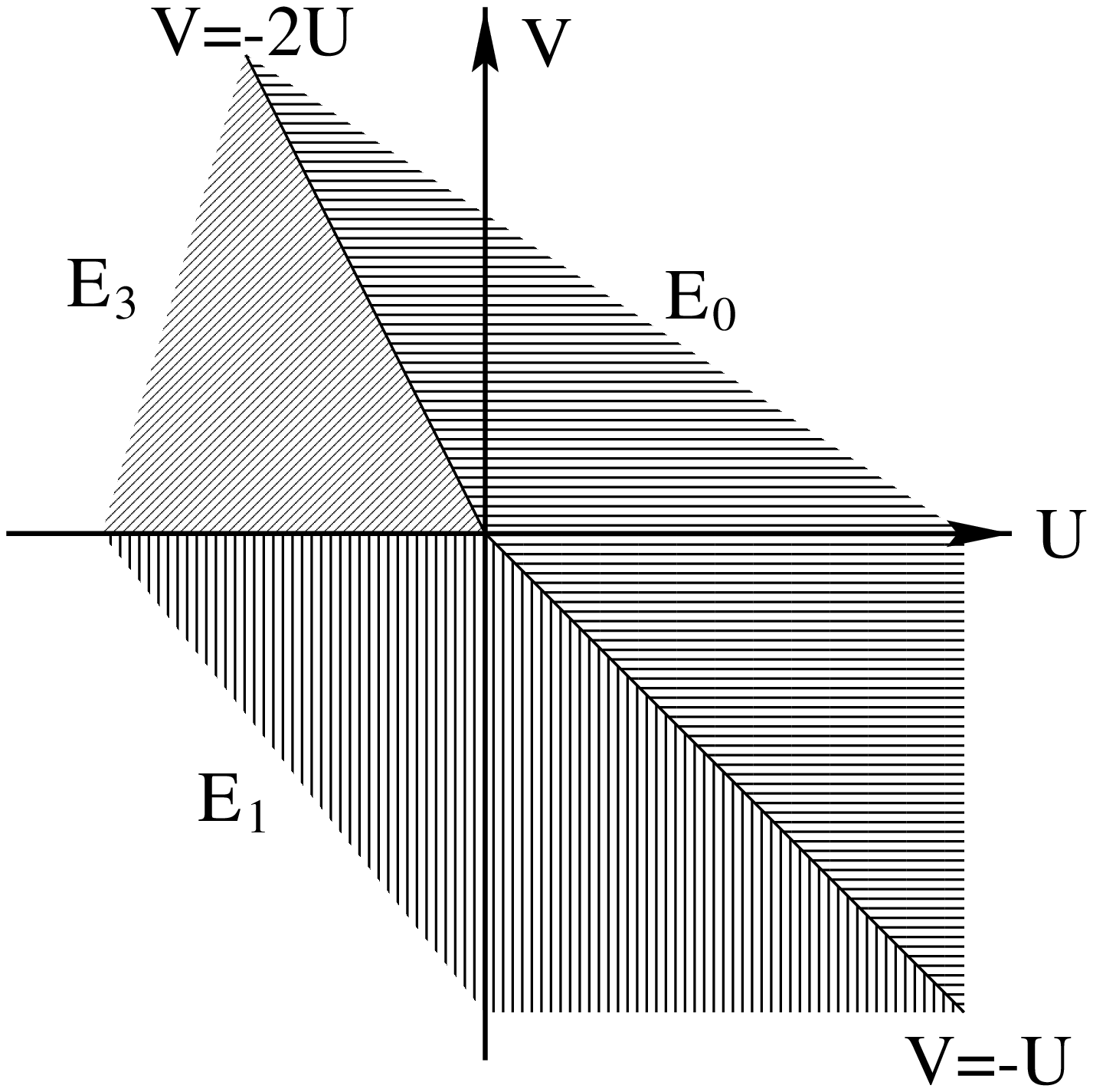}
}
\vskip .13 in
\caption{The half-filled strong-coupling $U-V$ phase diagram showing the
regions in which the single rung manifolds indicated have the lowest
energy.  As discussed in Section III, the $E_0$ manifold is a spin-gap
phase which has $d_{x^2-y^2}$-like power law pair field correlations
when it is doped.  The $E_3$ manifold is divided into a CDW phase and a
spin-gapped phase which exhibits $s$-wave like pairing correlations upon
doping.  The $E_1$ regime is an $SO(5)$ generalization of the Heisenberg
spin one chain.}
\end{figure}

\newpage

\begin{figure}[h]
\centerline{\epsfysize=5cm
\epsfbox{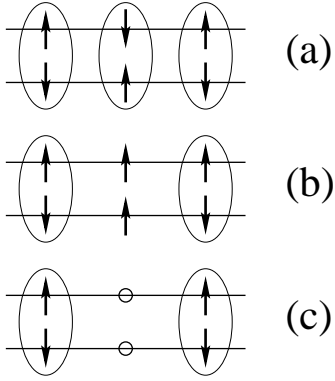}
}
\vskip .13 in
\caption{Schematic illustration of (a) the $E_0$ ground state, (b) a
spin $S=1$ magnon, (c) a hole pair. The ellipses signify a singlet
state.}
\end{figure}

\begin{figure}[h]
\centerline{\epsfysize=5cm
\epsfbox{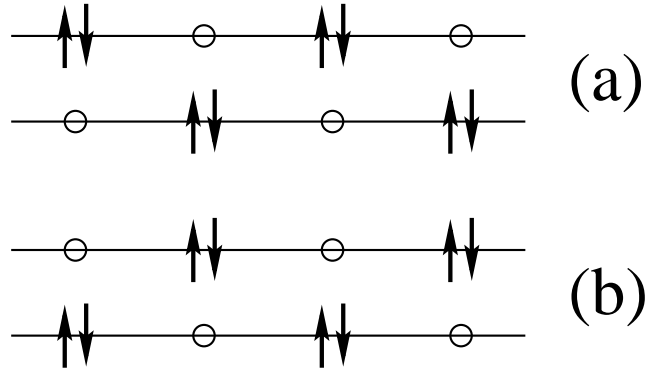}
}
\vskip .13 in
\caption{Illustration of the two-degenerate CDW $E_3$ ground states
which can occur when $K_3>h$.}
\end{figure}

\begin{figure}[h]
\centerline{\epsfysize=5cm
\epsfbox{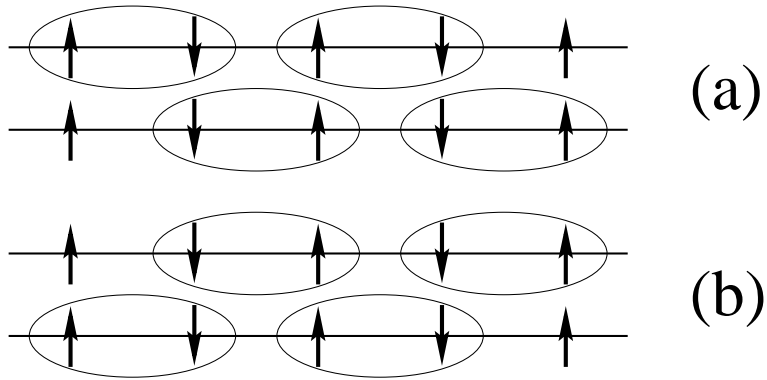}
}
\vskip .13 in
\caption{Schematic of two local configurations between which the system
resonates in the $E_1$ superspin phase.}
\end{figure}

\newpage

\begin{figure}[h]
\centerline{\epsfysize=8cm
\epsfbox{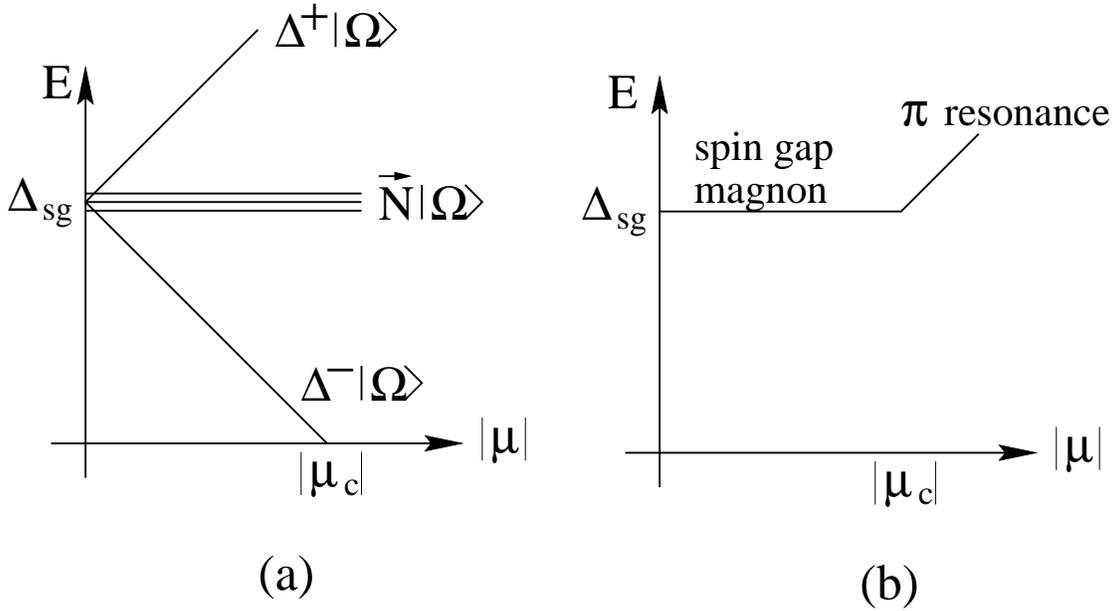}
}
\vskip .13 in
\caption{(a) The $SO(5)$ vector modes versus $\mu$. Here
$\Delta^+|\Omega\rangle$ and $\Delta^-|\Omega\rangle$ 
add or remove a pair, while $\vec N|\Omega\rangle$ represents 
the three $S=1$ magnon modes.  $\Delta_{sg} = J$ is the 
spin gap. (b) The $q=(\pi,\pi)$ magnon versus $\mu$.
These modes remain constant until $\mu$ exceeds 
$\mu_c=\Delta_{sg}/2$ and a finite pair hole density 
is formed.  When this happens, the energy of the
magnon-$\pi$ mode increases in energy as $2|\mu|$.
}
\end{figure}

\newpage

\begin{figure}[h]
\centerline{\epsfysize=8cm
\epsfbox{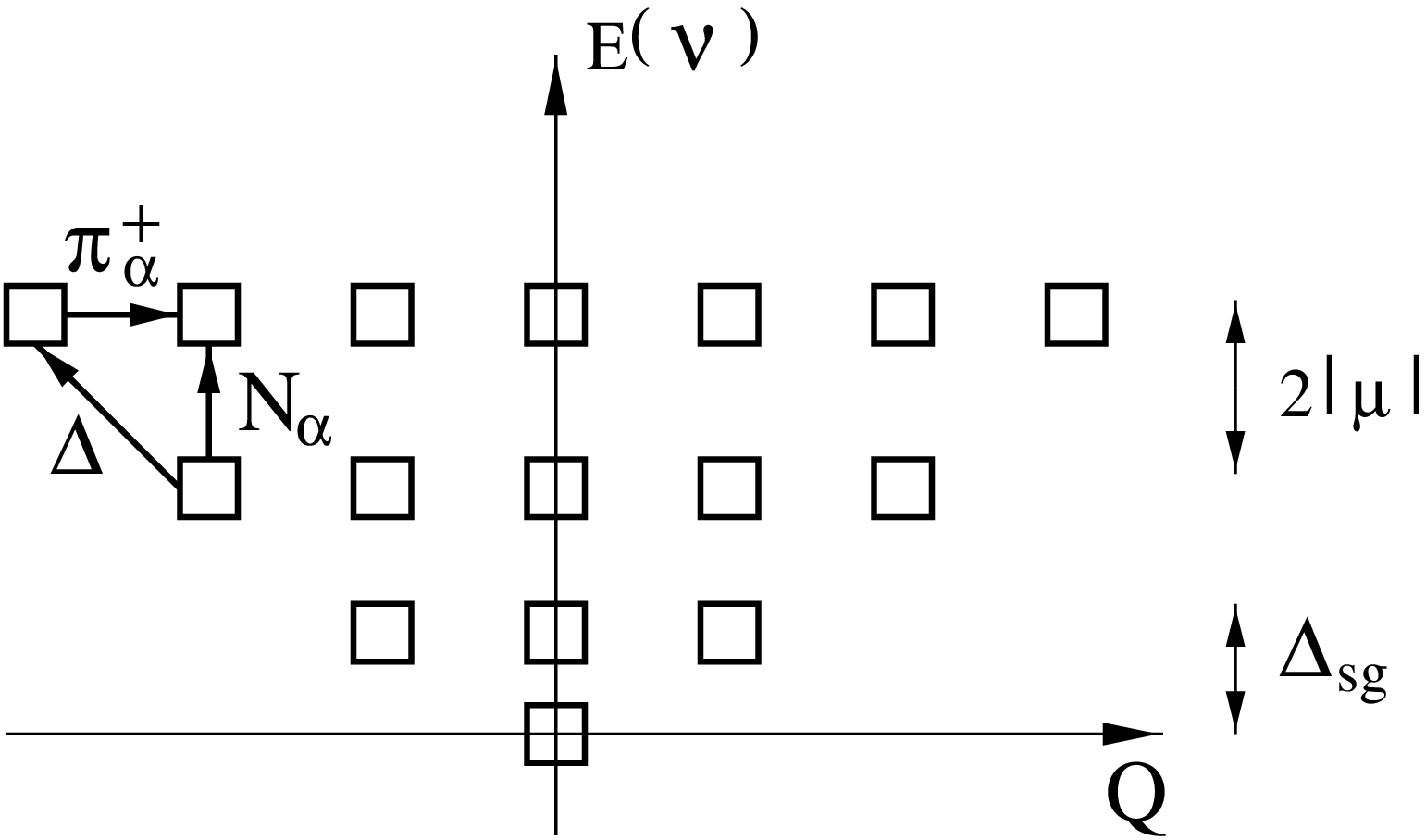}
}
\vskip .13 in
\caption{An $S_z=0$ slice of the $SO(5)$ pyramid. Here $Q=(N_e-N)/2$
is half of the electron charge measured from half-filling, and $\nu$
is related to the $SO(5)$ Casimir charge. Each box contains many 
states with the same $SO(5)$ quantum numbers. The ground states of 
a given charge sector lie within the edge of this triangle. One can
create a magnetic excitation of a charge $Q$ state either by the
$N_\alpha$ operator, or by first inserting a hole pair using the
$\Delta$ operator and then ``rotate" the resulting state by the
$\pi^\dagger_\alpha$ operator. This triangular relationship is 
indicated on the figure.}
\end{figure}


\begin{thebibliography}{10}

\bibitem{DR96} E.~Dagotto and T.M.~Rice, {\sl Science}, {\bf 271}, 618
(1996);  cond-mat/9509181.

\bibitem{NSW96} R.~Noack, D.J.~Scalapino, and S.R.~White, {\sl
Phil.~Mag.~B}, {\bf 74}, 485 (1996).

\bibitem{TTR96} M.~Troyer, H.~Tsunetsugu, and T.M.~Rice, {\sl
Phys.~Rev.~B}, {\bf 53}, 251 (1996); cond-mat/9510150.

\bibitem{BF96} L.~Balents and M.P.~Fisher, {\sl Phys.~Rev.~B}, {\bf 53},
12133 (1996);  cond-mat/9503045.

\bibitem{so5} S.C. Zhang,
{\em Science}, 275:1089, 1997.

\bibitem{demler} E.~Demler and S.C. Zhang,
{\em Phys. Rev. Lett.}, 75:4126, 1995.

\bibitem{rabello} S. Rabello, H. Kohno, E. Demler and S. C. Zhang
preprint, cond-mat/9707027.

\bibitem{henley} C. Henley, preperint, cond-mat/9707275.

\bibitem{burgess}
C.P. Burgess, J.M. Cline, R. MacKenzie and R. Ray, 
preprint, cond-mat/9707290.

\bibitem{meixner} S.~Meixner, W.~Hanke, E.~Demler and S.C. Zhang,
to be published in {\em Phys. Rev. Lett.}, 75, 1997; cond-mat/9701217.

\bibitem{eder} R.~Eder, W. Hanke, and S.C. Zhang,
preprint, cond-mat/9707233.

\bibitem{demler2} E.~Demler, H. Kohno and S.C. Zhang,
preprint, cond-mat/9710139.


\bibitem{shelton} D.G.~Shelton, D.~Sinichal, preprint, cond-mat/9710251.

\bibitem{lin} L.~Balents, M.~Fisher, and H.~Lin, to be published.

\bibitem{enrico} E.~Arrigoni and W.~Hanke, to be published. 

\bibitem{aklt} I.~Affleck, T.~Kennedy, E.H.~Lieb, and H.~Tasaki,
{\em Phys. Rev. Lett.}, 59:799, 1987.

\bibitem{Ref1} As discussed by Troyer et.~al \cite{TTR96}, the actual
spin gap evolves discontinuously upon doping due to the existance of
pair breaking quasi-particle (and bound states) $S=1$ excitations that
can be created in a doped ladder.  However, the spectral weight
associated with these quasi-partical pair excitations varies as the
doping and near half-filling is small compared to the collective nagnon
mode which we are discussing. 

\end{thebibliography}
\end{document}